\documentclass[pra,aps,twocolumn,superscriptaddress]{revtex4-1}
\usepackage{graphicx,color}
\usepackage{amsthm}
\usepackage{amsfonts}
\usepackage{algorithmic}
\usepackage{enumerate}
\usepackage{latexsym}
\usepackage{amsmath}
\usepackage{amssymb}
\usepackage[colorlinks=true,citecolor=blue,linkcolor=blue]{hyperref}

\usepackage{dcolumn}
\usepackage{bm}

\usepackage[T1]{fontenc}
\usepackage{mathptmx}

\begin{document}
\title{An intermediate phase induced by dilution in a correlated Dirac Fermi system}

\author{Lingyu Tian}
\affiliation{Department of Physics, Beijing Normal University, Beijing
100875, China}
\affiliation{Beijing Computational Science Research Center, Beijing
100193, China}
\author{Jingyao Meng}
\affiliation{Department of Physics, Beijing Normal University, Beijing
100875, China}
\affiliation{Beijing Computational Science Research Center, Beijing
100193, China}
\author{Tianxing Ma}
\email{txma@bnu.edu.cn}
\affiliation{Department of Physics, Beijing Normal University, Beijing
100875, China}

\begin{abstract}
Substituting magnetic ions with nonmagnetic ions is a new way to study dilution.
Using determinant quantum Monte Carlo calculations, we investigate an interacting Dirac fermion model with the on-site Coulomb repulsion being randomly zero on a fraction $x$ of sites.
Based on conductivity, density of states and antiferromagnetic structure factor, our results reveal a novel intermediate insulating phase induced by the competition between dilution and repulsion.
With increasing doping level of nonmagnetic ions, this nonmagnetic intermediate phase is found to emerge from the zero-temperature quantum critical point separating a metallic and a Mott insulating phase, whose robustness is proven over a wide range of interactions.
Under the premise of strongly correlated materials, we suggest that doping nonmagnetic ions can effectively convert the system back to the paramagnetic metallic phase. 
This result not only agrees with experiments on the effect of dilution on magnetic order but also provides a possible direction for studies 
focusing on the metal-insulator transition in honeycomb lattice like materials.
\end{abstract}


\maketitle
\section{Introduction}
Since a number of exotic phenomena are observed in graphene\cite{Novoselov2005,RevModPhys.81.109,RevModPhys.84.1067} and silicene\cite{PhysRevB.76.075131,PhysRevLett.108.155501,Houssa2015},
the honeycomb lattice has become another topic that has received much
theoretical\cite{PhysRevLett.120.116601,HUANG2019310} and experimental\cite{Cao2018,PhysRevResearch.3.033192} attention with respect to the square lattice.
In the absence of interactions, itinerant electrons on a honeycomb lattice form a Dirac spectrum, and the density of state (DOS) near Fermi energy, E=0, vanishes linearly, which is in strong contrast with the square lattice whose DOS diverges at E=0\cite{PhysRevB.101.205103}.
The difference is directly reflected in the onset of long-range antiferromagnetic (AFM) correlations at half filling: the AFM order can be exhibited in the square lattice for arbitrarily small on-site Coulomb repulsion $U$, whereas a finite value $U_{c}\thicksim4t$ is required in the honeycomb lattice\cite{PhysRevB.72.085123,Sorella2012}. In addition, a transition from a Dirac semimetal phase to an insulating phase is also found at $U_{c}$\cite{PhysRevX.3.031010,PhysRevB.103.155110}.
The massless Dirac fermions have advanced our understanding of physics beyond Landau's theory
of the Fermi liquid\cite{doi:10.1126/science.1222360,doi:10.1126/science.1248253}, which states that interacting metallic systems are similar to free Fermi systems.

In real materials, disorder is inevitably present and can be controlled by doping. Disorder plays an important role in many novel physical properties of modern science, touching upon topics from transport phase transition\cite{PhysRevLett.101.086401,PhysRevB.93.224203,PhysRevB.97.245114} and superconductivity\cite{Gantmakher2010,PhysRevB.90.094516} to quantum spin liquids\cite{PhysRevLett.118.107202,PhysRevX.8.041040,PhysRevB.98.134427,PhysRevLett.123.087201}.
However, different types of disorder may have opposite influences on the physical mechanism under the same model, which makes studies about disorder in correlated systems important and interesting.
For example, the nearest-neighbor hopping disorder is proven to enhance localization in the two-dimensional repulsion Hubbard model under half filling, while the site disorder reduces this effect. Interestingly, both types of disorder destroy the AFM order for dynamic properties\cite{PhysRevLett.87.146401}.
Another example is using strong-coupling perturbation theory to study the Anderson Hubbard model on the honeycomb lattice, where an intermediate metallic state is present between the Anderson insulator and Mott insulator under binary-alloy disorder but absent under uniformly distributed disorder\cite{PhysRevB.98.245105,PhysRevB.99.014204}.

Site dilution is the disorder achieved by substitution of magnetic ions with nonmagnetic ions.
It has been reported that doping nonmagnetic Pt and G into GaFe$_{4}$As$_3$ produces a different modification of the electronic structure and transition property\cite{WU2015342}.
In an insulating honeycomb magnet $\alpha-$RuCl$_{3}$, replacing the magnetic ions Ru$^{3+}$ with nonmagnetic ions Ir$^{3+}$ suppresses the magnetic order and induces a dilute quantum spin liquid state in the low-temperature region \cite{PhysRevLett.119.237203,PhysRevB.98.014407,PhysRevLett.124.047204}.
Many models have been used to reveal the phases in diluted systems \cite{PhysRevB.104.035116},and, motivated by cuprates such as La$_{2}$Cu$_{x}$Mg$_{1-x}$O$_{4}$, the site diluted Hubbard model with on-site repulsion $U$ being zero randomly on a fraction $x$ of sites has been investigated in two and quasi-two dimensions.
On the strong coupling square lattice, the AFM magnetic order at half filling disappears at $x_{c}$,  which is consistent with the classical percolation threshold $x_{c}^{\rm(perc,square)}$,
while on the Lieb lattice, $x_{c}$ is almost twice that of  $x_{c}^{(\rm perc,lieb)}$\cite{PhysRevB.101.165109}.
This difference emphasizes the central role of electron itinerancy in the magnetic response.
In related experiments, the honeycomb lattice provides a great platform to study dilution.
In CoTiO$_{3}$, a linear relation between dilution and the critical temperature of magnetic transition is observed over a wide dilution range\cite{Horsley_2022}.
Under out-of-plane interactions or second- and third-nearest-neighbor exchange interactions, the long-range AFM order survives well past the classical percolation threshold $x_{c}=0.3$\cite{PhysRevE.78.031136,Horsley_2022,PhysRevB.73.054422}.
Our study focuses on the magnetic phase transition caused by dilution in the system with on-site interaction, and it is also interesting to study whether magnetic order transition and metal-insulator transition take place concomitantly.

\begin{figure}[t]
\centerline {\includegraphics*[width=3.6in]{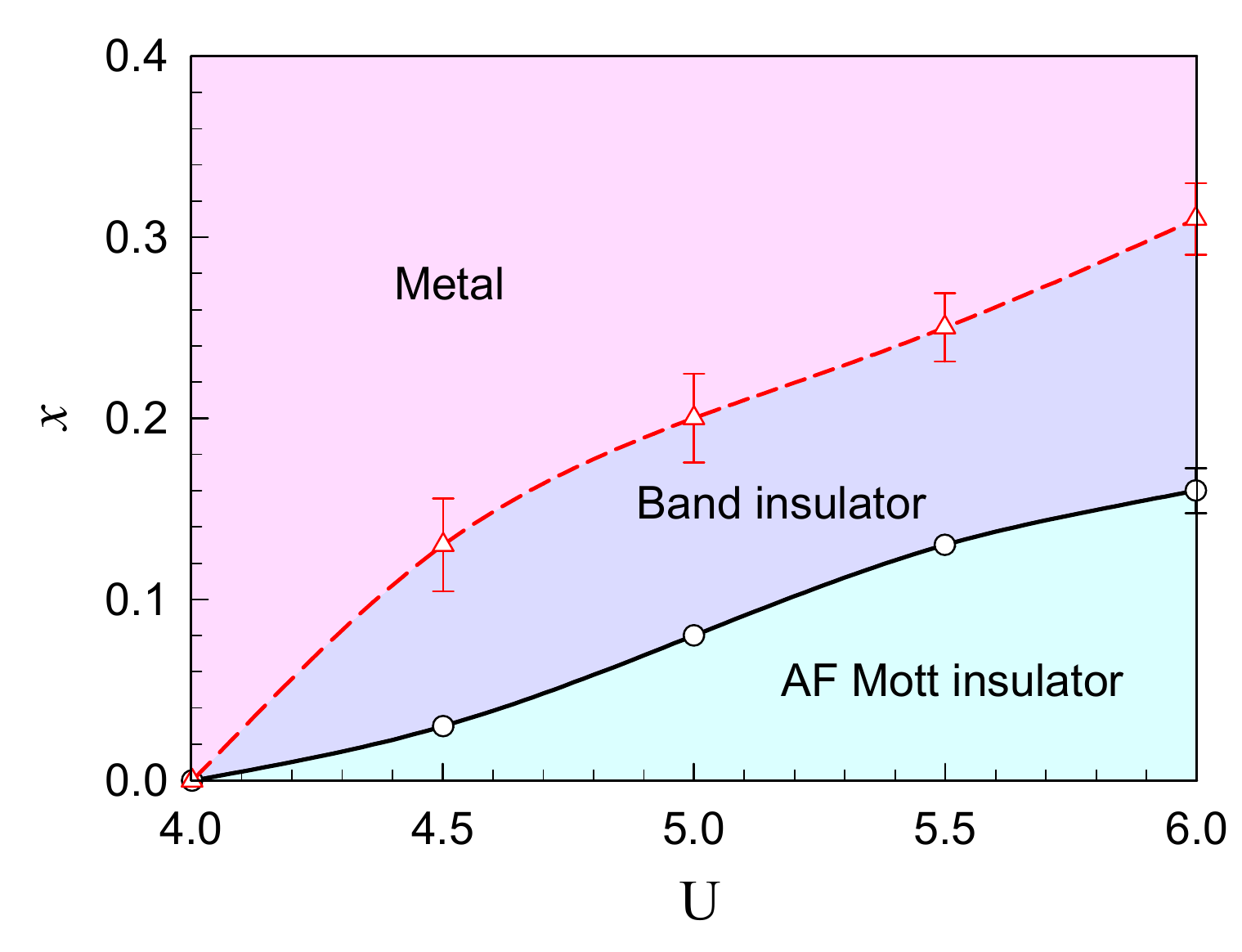}}
\caption{(Color online) Phase diagram of the Hubbard model on a $N=2L^{2}$ honeycomb lattice at half filling.
$x$ represents the percentage of the number of free sites with respect to the total number of lattice sites, and $U$ labels the on-site Coulomb repulsive interaction. Phase boundaries are determined by the conductivity, density of states at the Fermi energy and finite size scaling of the AFM spin structure factor. There exists an intermediate phase $--$ band insulator $--$ between the AFM Mott insulator and metal. These phases are labeled by different colors in the figure: AFM Mott insulator (green), band insulator (blue), metal (pink).}
\label{Fig:phase}
\end{figure}

Here, we use determinant quantum Monte Carlo (DQMC) simulations to examine the effect of dilution on the ground state properties of honeycomb lattices, including transport and magnetic properties, of the half-filled case.
Our key result is that an intermediate gapped insulating phase without magnetic order is identified between the AFM Mott insulator and the metal
and is robust over a wide range of $U$, which is summarized in the phase diagram in Fig. \ref{Fig:phase}.
The red dashed line represents the phase boundary between the gapped insulator and metal, which is determined by the conductivity and density of states.
The black solid line indicates an AFM phase transition confirmed by the spin structure factor.
Our results have the possibility to be realized in optical lattice experiments and were previously demonstrated in a one-dimensional optical lattice to induce multiple phase transitions by introducing randomness into the interaction distribution via Feshbach resonances\cite{PhysRevLett.95.170401,PhysRevA.95.021601}.

\section{Model and method}
We consider a modified version of the Hubbard model with site-dependent repulsion, described by the Hamiltonian:
\begin{eqnarray}
\label{Hamiltonian}
\hat H & = & -t \sum_{{\bf i}\in A,{\bf j}\in B,\sigma}(\hat c_{{\bf i}\sigma}^\dagger \hat c_{{\bf j} \sigma}+H.c.)^{\phantom{\dagger}}-\mu \sum_{{\bf i}\sigma} \hat n_{{\bf i}\sigma} \nonumber\\
          && +\sum_{{\bf i}}U_{i}(\hat n_{{\bf i}\uparrow}-\frac{1}{2})(\hat n_{{\bf i}\downarrow}-\frac{1}{2}).
\end{eqnarray}
Here, $\hat c_{{\bf i}\sigma}^\dagger$ $(\hat c_{{\bf j}})$ indicates creation (annihilation) electron operators in second-quantized formalism, and $\hat n_{{\bf i} \sigma}=\hat c_{{\bf i}\sigma}^\dagger \hat c_{{\bf i}\sigma}$ is the occupancy number operator. The first term on the right-hand side of Eq. (\ref{Hamiltonian}) denotes in-plane hopping between nearest neighbors, and in our paper, the hopping amplitude is set as $t=1$, thus defining the energy scale. The last term includes the chemical potential $\mu$, and we set $\mu=0$, a choice that makes the studied system precisely half-filled and protects particle-hole symmetry.

We introduce dilution by allowing for random distribution of the site-dependent Coulomb repulsion $U_{i}$ in the second term, such that the on-site interaction on a fraction x of the sites is suppressed:

$$U_{i}=\left\{
\begin{aligned}
U  &  & {1-x} \\
0  &  & {x}
\end{aligned}
\right.
$$
This type of disorder is generated in a canonical ensemble, i.e., we have a fraction $Nx$ of sites with $U=0$ for a given concentration x, where N is the total number of sites, so that there are no charge fluctuations on the free sites. Here, we consider a 2$L^{2}$ honeycomb lattice with a linear size of $L=12$.
For dilution concentration $x$, where $Nx$ is not an integer number, we calculate a weighted average of its adjacent integers.
Our results are obtained by averaging over 20 disorder realizations.

We probe the transport and magnetic properties of the half-filled diluted honeycomb model by means of DQMC simulations.
In this method, the Hamiltonian is mapped onto free fermions moving in a fluctuating space- and imaginary time-dependent auxiliary field by the Hubbard-Stratonovich (HS) transform\cite{PhysRevD.24.2278,PhysRevB.28.4059}.
This HS field is initialized randomly, and a local flip is attempted with the acceptance rate determined by the Metropolis algorithm.
A QMC sweep is completed when the process of changing the auxiliary field variable traverses the entire space-time.
In our simulations, 4,000 warm-up sweeps were used to equilibrate the system, and then 48,000 sweeps were conducted for measurements.
The number of measurements was split into 10 bins, which provide the basis of coarse-grain averages and errors  estimated based on standard deviations from the averages.
The errors from the Suzuki-Trotter decomposition are proportional to $(\Delta\tau)^{2}$, where $\Delta\tau=\beta/M$ is the imaginary-time interval, so we set $\Delta\tau=0.1$ to guarantee that the systematic errors are smaller than those associated
with statistical sampling\cite{PhysRevB.85.125127}.
As with many fermionic QMC methods, the DQMC method also suffers from the minus-sign problem; however, the particle-hole symmetry makes our system free of the sign problem so that the simulation can be performed at low enough temperature to converge to the ground state.

With the aim of exploring the phase transitions between metal and insulating phases, we compute the $T$-dependent direct-current conductivity:

\begin{equation}
 \sigma_{dc}(T) =
   \frac{\beta^2}{\pi} \Lambda_{xx} ({\bf q}=0,\tau=\beta/2).
\label{eq:dc}
\end{equation}
where $\beta = 1/T$ is the inverse temperature and the momentum q- and imaginary time $t$-dependent current-current correlation functions $\Lambda_{xx} ({\bf q},\tau)$ are expressed as $\Lambda_{xx} ({\bf q},\tau)$ = $\langle \hat j_{x} ({\bf q},\tau)\hat j_{x} (-{\bf q},0)\rangle$.
$\hat j_{x} ({\bf q},\tau)$ is the Fourier transform of the $\tau$-dependent current density operator in the x direction.
This approximation has been extensively employed to identify metal-insulator transitions for either disordered or clean systems\cite{PhysRevB.54.R3756,doi:10.1126/science.aau7063}. To establish the existence of the Mott insulator, we define $N(0)$, the density of states at the Fermi energy, as

\begin{equation}
N(0) \simeq \beta \times G({\bf r}=0, \tau=\beta/2).
\label{eq:dos}
\end{equation}
Here, $G$ is the real-space single-particle Green function\cite{PhysRevLett.75.312,PhysRevB.97.085123}.

In addition to transport properties, we also examine the magnetic properties by investigating the antiferromagnetic structure factor\cite{Meng2010}:
\begin{equation}
S_{AFM}=\frac{1}{N}\langle \langle (\sum_{{\bf r}\in A} \hat S_{{\bf r}}^{z}-\sum_{{\bf r}\in B} \hat S_{{\bf r}}^{z})^2 \rangle \rangle.
\label{eq:Saf}
\end{equation}
where $S_{{\bf r}}^{z}$ represents the $z$ component spin structure factor operator on the A/B sublattices of the honeycomb lattice.

\section{Results and Discussion}

\begin{figure}[t]
\centerline {\includegraphics*[width=3.6in]{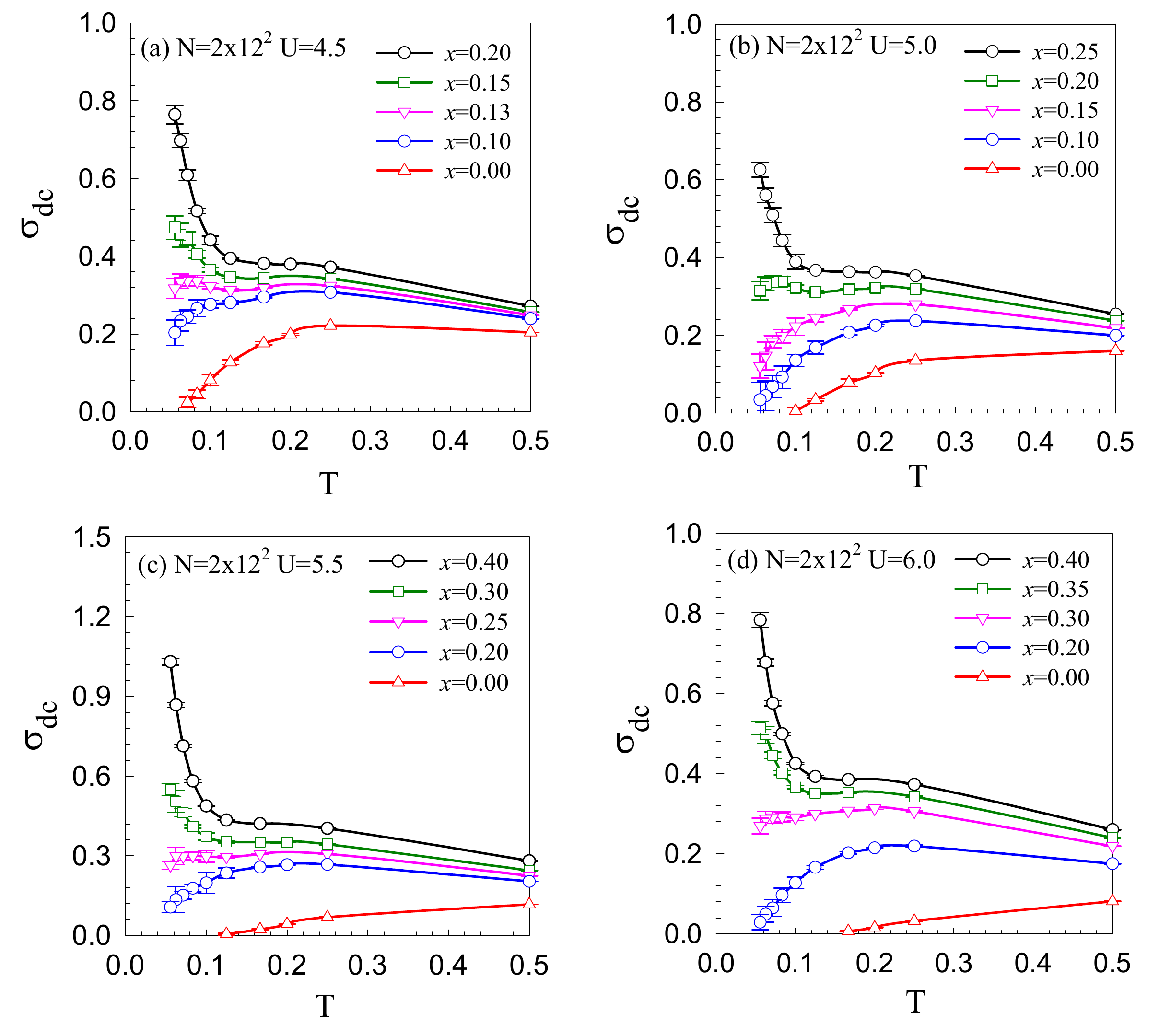}}
\caption{(Color online) The conductivity is shown as a function of temperature for various dilution concentrations at (a) $U=4.5$, (b) $U=5.0$, (c) $U=5.5$, and (d) $U=6.0$. Differences in the conductivity behavior indicate a dilution-driven insulator-metal transition. Data points are averages over $20$ disorder realizations.}
\label{Fig:dc}
\end{figure}

As mentioned above, numerous studies have confirmed that the honeycomb lattice is in a Mott insulating phase at $U\geq4t$\cite{PhysRevB.72.085123}. When the system is diluted, the appearance of free lattices weakens the repulsive potential, making it impossible to maintain the half-filled system in an insulating state.
Thus, we expect a critical point of dilution, which represents an insulator-metal transition.
This can be checked by the conductivity calculated through Eq. (\ref{eq:dc}).
Figure \ref{Fig:dc} shows the temperature $T$ dependence of conductivity $\sigma_{dc}$ at $U\geq4.5$ for different dilution concentrations $x$.
Regardless of $x$, $\sigma_{dc}$ increases as $T$ decreases for $T\geq0.25$. When $T$ continues to decrease, the conductivity curves behave differently:
at $x=0.00,0.10$ shown in Fig.\ref{Fig:dc} (a), $\sigma_{dc}$ decreases as the temperature decreases, even approaching the origin as $T\rightarrow0$, which is regarded as an insulating state; conversely, at $x=0.15,0.20$, $\sigma_{dc}$ continues to increase as the temperature decreases, which is regarded as a metallic state.
The behavior of $d\sigma_{dc}/dT$ changing from positive to negative indicates an insulator-metal transition (IMT).
The point at which conductivity is independent of temperature is considered the critical IMT point of dilution $x_{c}^{IMT}$, and we can roughly estimate $x_{c}^{IMT}$ for various interactions from Fig. \ref{Fig:dc}: $x_{c}^{IMT}\sim0.13$ for $U=4.5$; $x_{c}^{IMT}\sim0.20$ for $U=5.0$; $x_{c}^{IMT}\sim0.25$ for $U=5.5$; and $x_{c}^{IMT}\sim0.31$ for $U=6.0$.
The critical value becomes larger as the interaction strength increases.

\begin{figure}[t]
\centerline {\includegraphics*[width=3.6in]{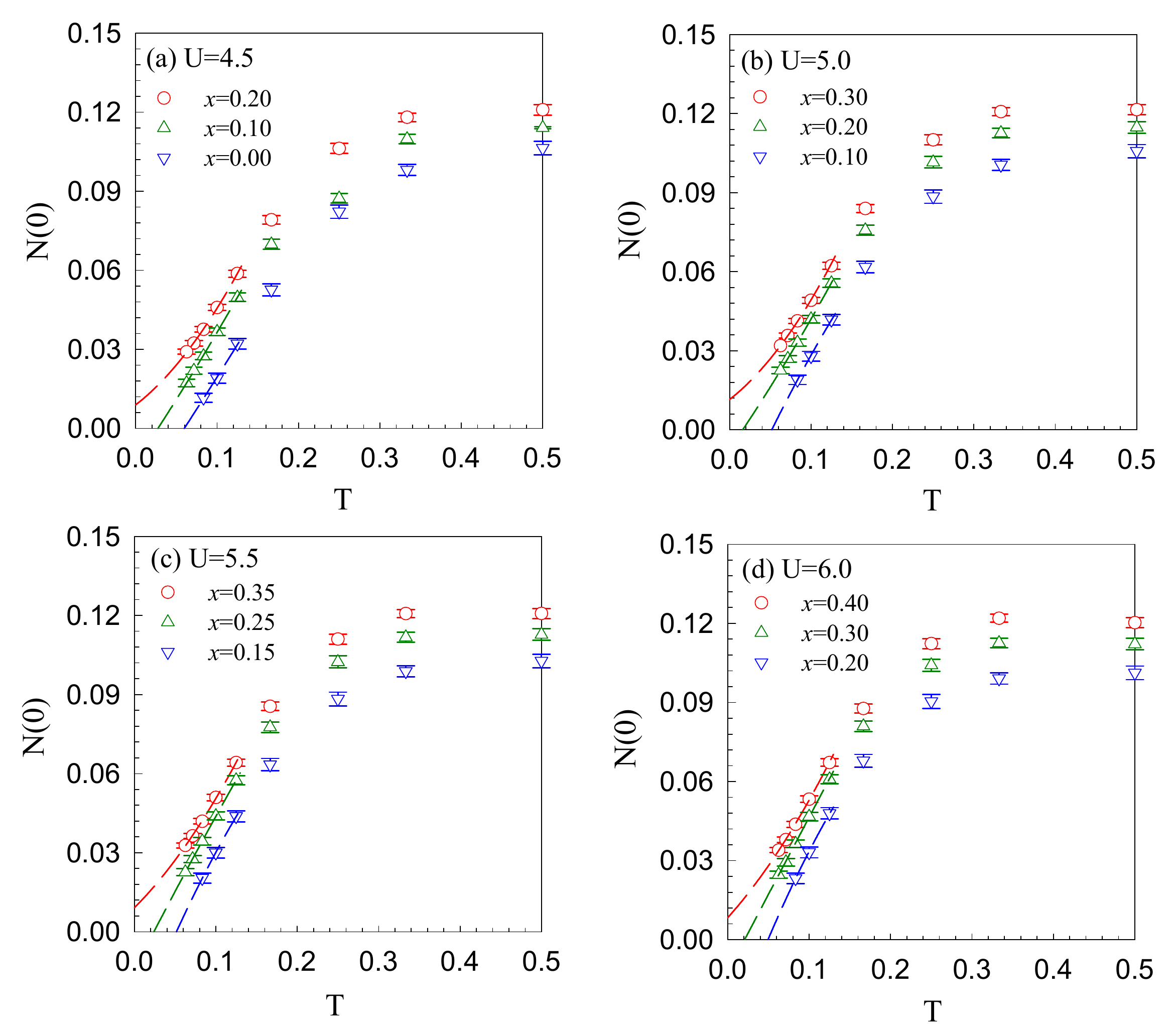}}
\caption{(Color online) The density of states at the Fermi energy $N(0)$ computed as a function of temperature for various dilution concentrations at (a) $U=4.5$, (b) $U=5.0$, (c) $U=5.5$, and (d) $U=6.0$. When increasing the dilution concentration, $N(0)$ approaches a finite value at $T\rightarrow0$, which indicates that the system is a gapped system.
The measurement of $N(0)$ once again verifies the conclusion that dilution induces insulator-metal transition.}
\label{Fig:dos}
\end{figure}

To further support our analysis of the insulator-metal transition, we calculated the density of states at the Fermi energy $N(0)$ for system behavior in metallic or insulating phases.
In Fig.\ref{Fig:dos} (a), $N(0)$ approaches zero at $T\rightarrow0$ in the case of $x\leq0.1$, suggesting that the system opens a gap.
As $x$ increases, $N(0)$ in the thermodynamic limit gradually becomes a finite value, indicating that the gap closes\cite{PhysRevB.104.045138}. The result of $N(0)$ is consistent with that measured by conductivity; that is, an increased dilution concentration causes an insulator to become a metallic phase by closing the energy gap.
The insulating phases are always gapped, as evidenced by the phenomenon that $N(0)$ converges to zero, and are divided into two types by the following magnetic calculations.

\begin{figure}[t]
\centerline {\includegraphics*[width=3.6in]{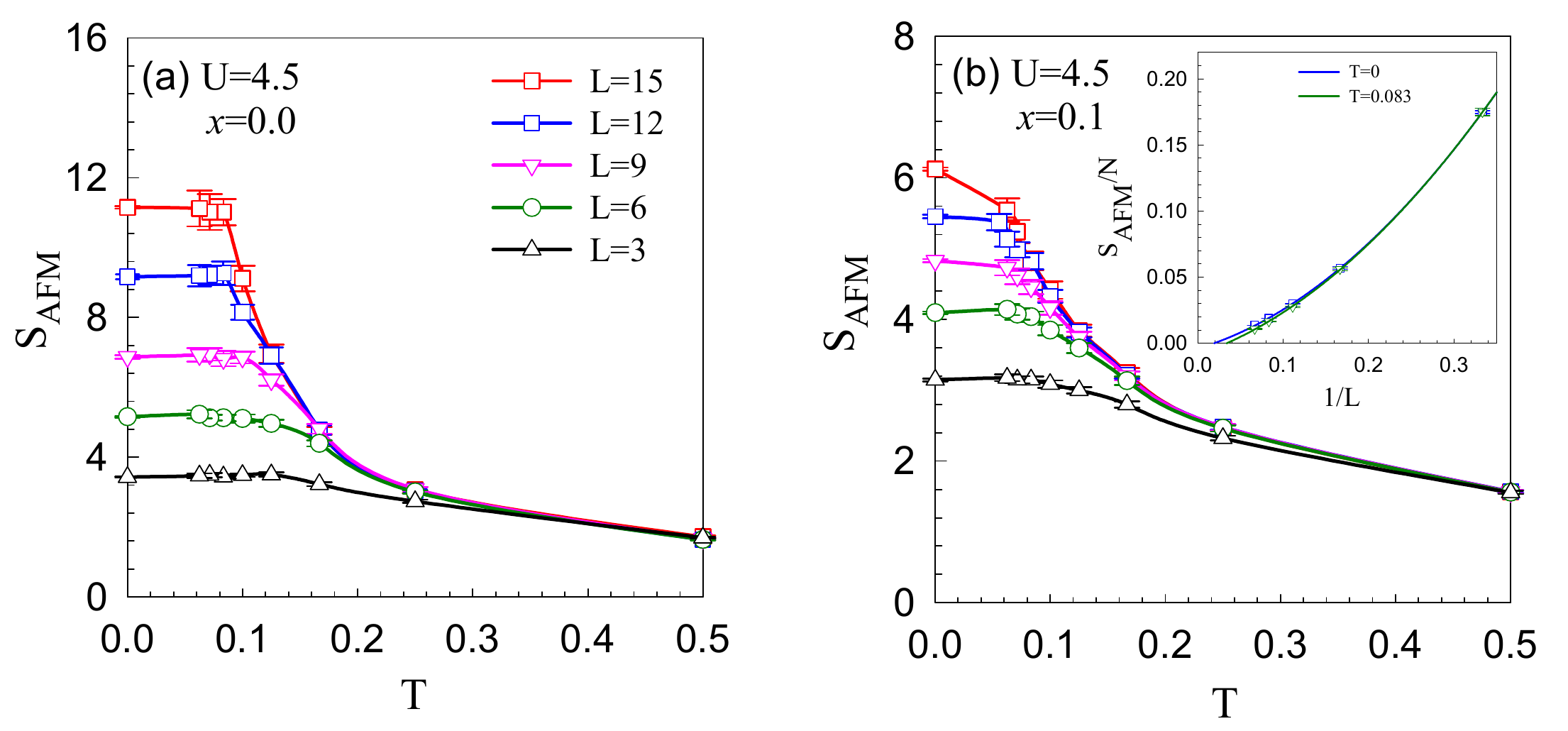}}
\caption{(Color online) Temperature dependence of the AFM spin structure factor $S_{AFM}$ with different lattice sizes for (a) $U=4.5, x=0.0$ and (b) $U=4.5, x=0.1$. Inset: finite size extrapolation of the AFM spin structure factor with different temperatures $T=0$ and $T=0.083$ for $U=4.5, x=0.1$. The agreement of the qualitative results for $T=0$ and $T\approx0.083$ validates reliability for studying the ground state magnetism by using $T\approx0.083$ data.}
\label{Fig:afm}
\end{figure}

\begin{figure}[t]
\centerline {\includegraphics*[width=3.6in]{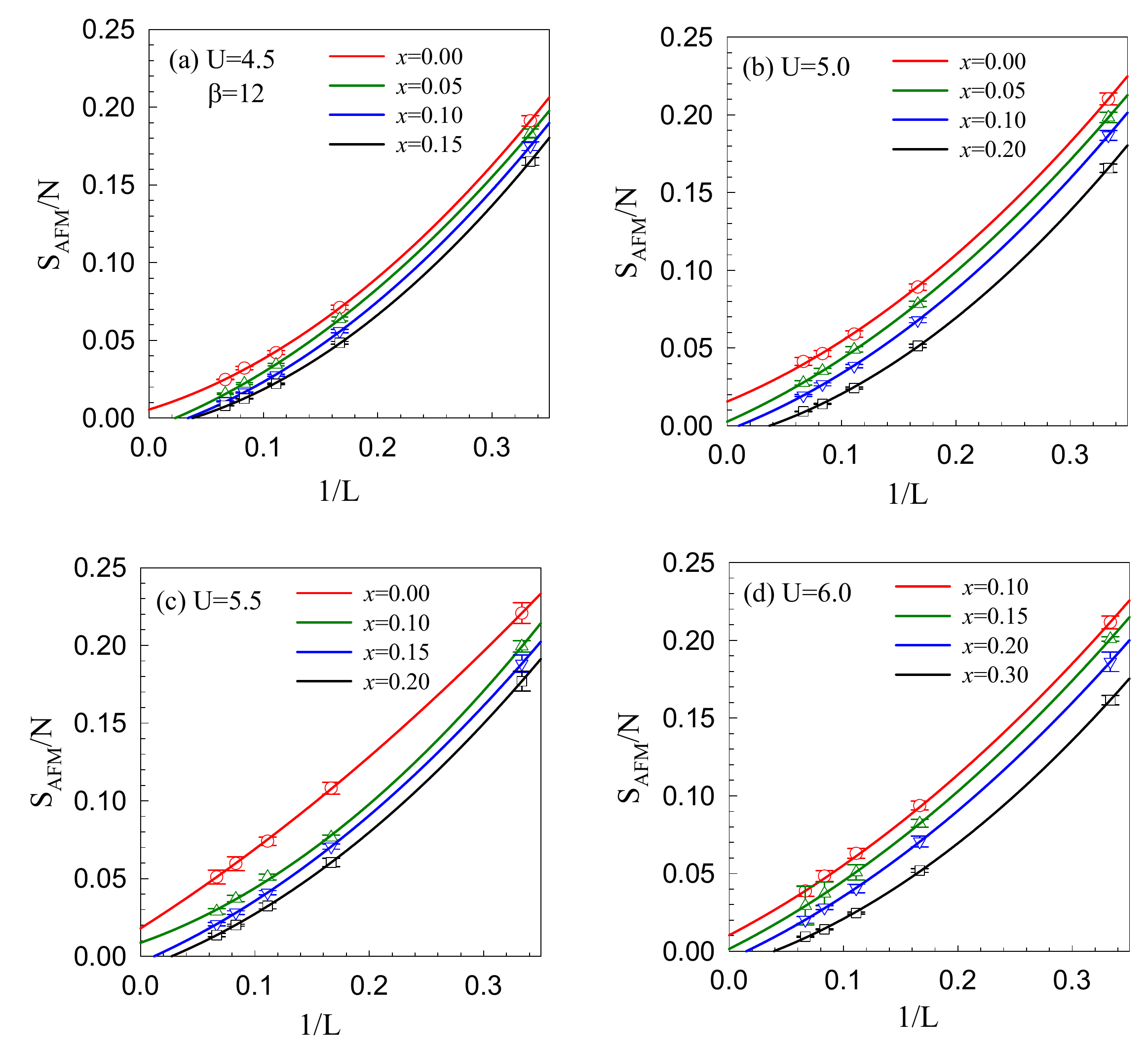}}
\caption{(Color online) The normalized AFM spin structure factor $S_{AFM}/N$ plotted as a function of $1/L$ for different dilution concentrations at fixed interactions: (a) $U=4.5$, (b) $U=5.0$, (c) $U=5.5$, and (d) $U=6.0$. Solid lines are obtained by fitting a second-order polynomial to the DQMC data. A finite y-axis intercept in the thermodynamic limit represents the existence of long-range antiferromagnetic order.}
\label{Fig:scaling}
\end{figure}

Another physical property of interest is the magnetic order, which has been shown to be suppressed by dilution\cite{PhysRevB.101.165109}.
Fig.\ref{Fig:afm} (a) and (b) summarize the relationship between the AFM spin structure factor $S_{AFM}$ and temperature for $x=0.0$ and $x=0.1$.
In each case, $S_{AFM}$ develops as the lattice size $L$ increases and the temperature decreases.
When reaching a lattice-dependent temperature of approximately $T\approx0.083$, $S_{AFM}$ saturates, which is more obvious in the undiluted case.
We compare the finite size scaling results for $T=0$ and $T\approx0.083$ in the inset of Fig.\ref{Fig:afm} (b), in which the result of $T=0$ is obtained from the ALF code.
Although there is a slight difference in the value of $S_{AFM}$ between these two temperatures shown in Figure (b), the intercept with the vertical axis ($1/L = 0$) is always negative, and the qualitative results are the same.
Considering the temperature effect of antiferromagnetism, we hypothesize that the result of $T\approx0.083$ can be regarded as the result for the ground state.

Following the procedure adopted for the insert, in Fig.\ref{Fig:scaling} we extrapolate the results of $T\approx0.083$ to $L\rightarrow\infty$, which is small enough to capture the ground-state magnetization and avoids unnecessarily complicated large simulations.
It is known that $\lim_{N\rightarrow\infty}(S_{AFM}/N)>0$ indicates the onset of long-range AFM order\cite{PhysRevB.104.035104}.
As shown in Fig.\ref{Fig:scaling} (a), the value of $S_{AFM}(1/L\rightarrow0)$ changes from positive to negative as $x$ increases to 0.05 at $U=4.5$, which demonstrates that the $T=0$ system undergoes a transition from magnetic order to magnetic disorder in the thermodynamic limit.
Similarly, the critical values of the AFM phase transition $x_{c}^{AFM}$ can be obtained from Fig.\ref{Fig:scaling} (b) $-$ (d): $x_{c}^{AFM}=0.05\sim0.10$ for $U=5.0$, $x_{c}^{AFM}=0.10\sim0.15$ for $U=5.5$, and $x_{c}^{AFM}=0.15$ for $U=6.0$.
The dilution suppresses the long-range AFM order, while $U$ plays the opposite role.
It is worth noting that for each interaction strength, $x_{c}^{AFM}$ is smaller than $x_{c}^{IMT}$, and this difference is too large to be explained by errors.
Thus, there is an intermediate phase existing between the AFM Mott insulating phase and metallic phase.
This intermediate phase is denoted as a band insulating phase due to its insulativity, energy gap and lack of magnetic order.
We summarize these results as a two-dimensional phase diagram shown in Fig.\ref{Fig:phase}.

\begin{figure}[t]
\centerline {\includegraphics*[width=3.6in]{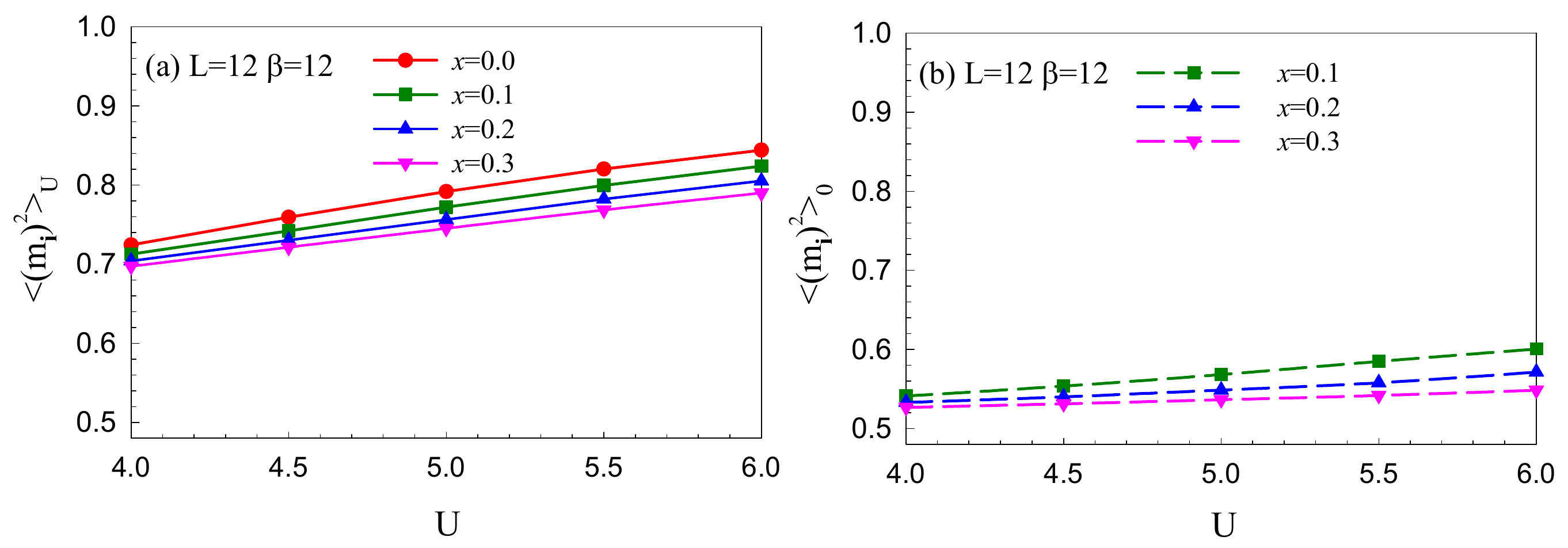}}
\caption{(Color online) Average local moment within repulsive (solid line) and free (dashed line) ($U=0$) sites as a function of repulsion strength for different dilution concentrations at temperature $T\approx0.083$. The data of the local moment inversely illustrate the double occupancy behavior.}
\label{Fig:m}
\end{figure}

Finally, we discuss the effect of dilution on the local moment defined as $\langle(\hat m_{i}^{z})^{2}\rangle \equiv \langle(
\hat n_{{\bf i} \uparrow} - \hat n_{{\bf i} \downarrow})^{2} \rangle$.
In the clean system, the local moment increases monotonically with increasing $U$\cite{PhysRevB.31.4403}.
At half filling, at $U=0$, both up spin and down spin occupy $0.5$, resulting in $\langle(\hat m_{i}^{z})^{2}\rangle=1-2*d=0.5$. When $U$ approaches infinity, all sites are singly occupied so that $\langle(\hat m_{i}^{z})^{2}\rangle=1$, corresponding to the spin-$\frac{1}{2}$ Heisenberg case.
Figure \ref{Fig:m} shows the dependence of the average local moment for repulsive $\langle(\hat m_{i}^{z})^{2}\rangle_{U}$ and free sites $\langle(\hat m_{i}^{z})^{2}\rangle_{0}$ on interaction.
For the former, $U$ has a positive role on $\langle(\hat m_{i}^{z})^{2}\rangle_{U}$; conversely, dilution has a negative effect.
The competition with the interaction also reflects that the insulator-metal transition may be driven by dilution.
In Figure (b), $\langle(\hat m_{i}^{z})^{2}\rangle_{0}$ increases with $U$ under each $x$, which is caused by the leaking of interaction into the free sites.
In the range of $U<6$, our results are consistent with Ref.\cite{PhysRevB.95.075142}, and when $U$ increases to a large value, the repulsive sites push electrons to the free sites and induce them to be doubly occupied, so we speculate that $\langle(\hat m_{i}^{z})^{2}\rangle_{0}$ will finally taper off to 0.5\cite{PhysRevB.95.075142}.

\begin{figure}[t]
\centerline {\includegraphics*[width=3.6in]{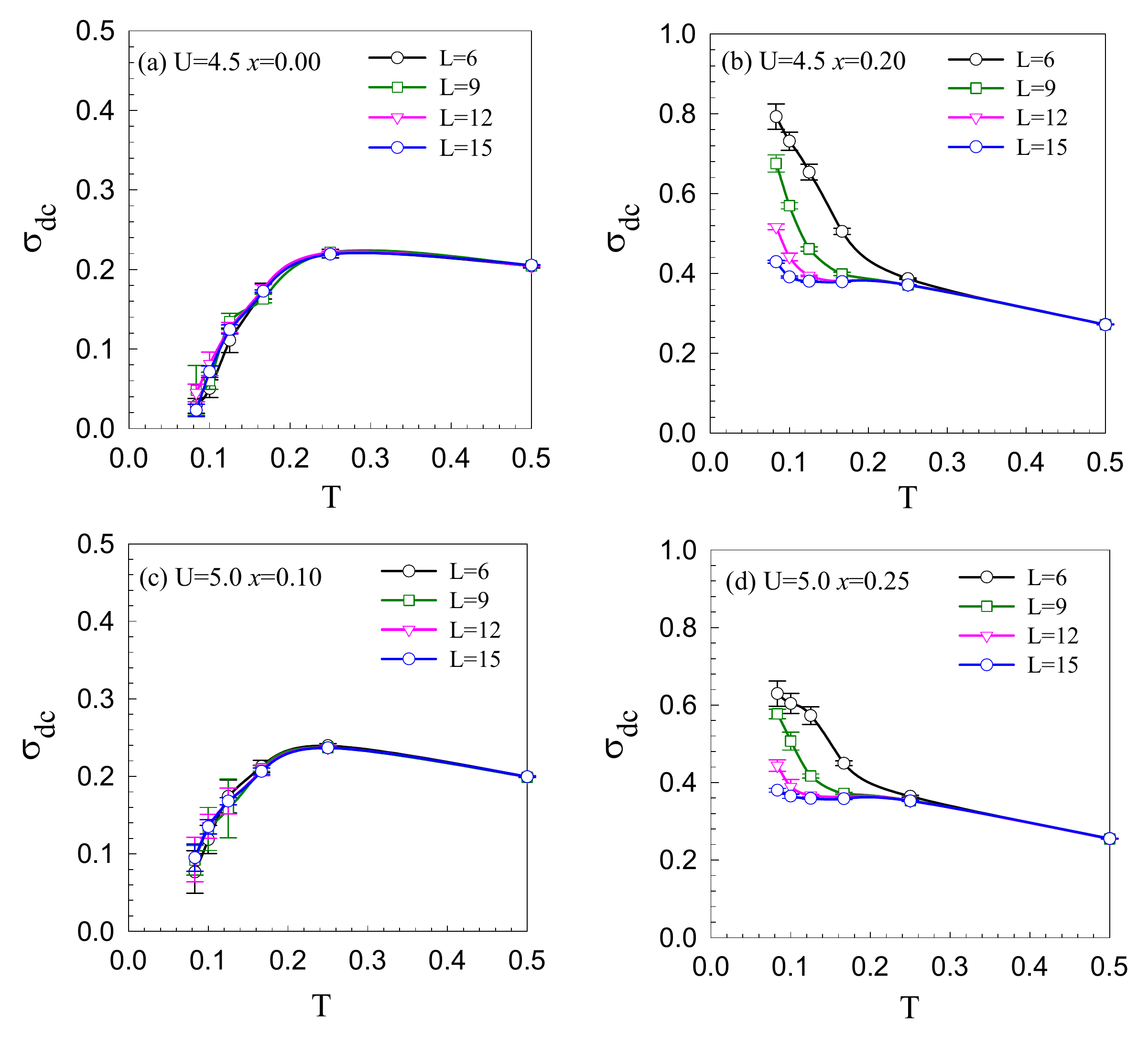}}
\caption{(Color online) The conductivity dependence on temperature for various lattice sizes at (a) $U=4.5$, $x=0.0$, (b) $U=4.5$, $x=0.20$, (c) $U=4.5$, $x=0.10$, and (a) $U=4.5$, $x=0.25$. In these figures, the systems of (a) and (c) are the insulating state, and those of (b) and (d) are the metallic state.}
\label{Fig:uniform}
\end{figure}

\section{Conclusions}

In conclusion, we use DQMC simulations to study the ground state properties of honeycomb lattices in the Hubbard model, with the on-site Coulomb repulsive interaction being deactivated randomly for a fraction $Nx$ of sites.
We employed the conductivity to examine the insulator-metal transition for $U=4.5,5.0,5.5,6.0$ and found that the critical value of IMT $x_{c}^{IMT}$ monotonously increases with the studied $U$.
We also use the density of states at Fermi energy to verify that the insulating phase is always gapped.
We calculated the AFM spin factor to investigate the influence of dilution on magnetic order, and the long-range AFM order vanishes as the dilution concentration increases at a fixed repulsion.
Because $x_{c}^{AFM}$ is lower than $x_{c}^{IMT}$ for each repulsion, we suggest that a new band-insulating intermediate phase appears between the AFM Mott insulator and metal, in which the energy gap is present but the magnetic order is absent.

Magnetic dilution has been the subject of extensive research efforts in the context of percolation phenomena.
The effect of dilution on magnetic order has received extensive attention for graphene-like materials\cite{Horsley_2022,Hallas2019}.
In this paper, we focus on a Dirac fermion system with on-site interactions and study the effects of dilution on both the magnetic order and insulator-metal transition.
Our results facilitate experimental studies related to honeycomb lattices, where phase transitions can be induced by doping nonmagnetic ions.

The relationship between $x_{c}^{AFM}$ and the percolation threshold has been discussed in previous studies\cite{PhysRevB.62.6629,PhysRevB.98.165142}.
Although all values of $x_{c}^{AFM}$ in our paper are smaller than the percolation threshold for the honeycomb lattice,
$x_{c}^{AFM}$ shows a clear growth trend with increasing repulsion.
Ref.\cite{PhysRevB.106.075146} found that
the long-range AFM order depends on the existence of local moments at the $U=0$ sites.
At fixed concentrations, $\langle(\hat m_{i}^{z})^{2}\rangle_{0}$ increases with $U$ (see Figure \ref{Fig:m}), and it can be deduced that $\langle(\hat m_{i}^{z})^{2}\rangle_{0}$ will approach the percolation threshold when $U$ increases to a certain extent.

\section{APPENDIX}

\setcounter{equation}{0}
\setcounter{figure}{0}
\renewcommand{\theequation}{A\arabic{equation}}
\renewcommand{\thefigure}{A\arabic{figure}}
\renewcommand{\thesubsection}{A\arabic{subsection}}
\
To make our phase diagram more convincing, the order parameters calculated on a finite lattice must all be extrapolated to the thermodynamic limit.
In the main text, we have demonstrated the finite size effect of the AFM spin structure factor and density of states. Here, we discuss the finite size effect of conductivity.
In Fig. \ref{Fig:uniform}, we plot $\sigma_{dc}$ as a function of $T$ for different sizes $L=6,9,12,15$. Figures (a) and (c) show that the lattice size has little effect on $\sigma_{dc}$, and
the behaviors of conductivity on all studied sizes are indicative of an insulating state.
In Figures (b) and (d), although there is a negative correlation between conductivity and lattice size, there is no qualitative change in their behavior, which always reflects the characteristics of the metallic phase.
Moreover, as the lattice size increases, the size effect on conductivity weakens.

The above results are consistent with the consensus: the size effect is larger for metallic systems than for gapped systems.
The results in Figure \ref{Fig:uniform} guarantee the accuracy of the phase diagram.

\noindent
\underline{\it Acknowledgments} ---
This work is supported by NSFC (No. 11974049). The numerical simulations were performed at the HSCC of Beijing Normal University and on the Tianhe-2JK in the Beijing Computational Science Research Center.

\bibliography{Reference}

\end{document}